\documentclass[sn-mathphys,Numbered]{sn-jnl}

\usepackage{graphicx}%
\usepackage{multirow}%
\usepackage{amsmath,amssymb,amsfonts}%
\usepackage{amsthm}%
\usepackage{mathrsfs}%
\usepackage[title]{appendix}%
\usepackage{xcolor}%
\usepackage{textcomp}%
\usepackage{manyfoot}%
\usepackage{booktabs}%
\usepackage{algorithm}%
\usepackage{algorithmicx}%
\usepackage{algpseudocode}%
\usepackage{listings}%
\usepackage{float}
\usepackage{natbib}
\usepackage{bm}
\usepackage{rotating}
\usepackage{dcolumn}
\usepackage{mathtools}
\usepackage[caption=false]{subfig}
\usepackage{empheq}
\usepackage{makeidx}
\usepackage{bookmark}

\theoremstyle{thmstyleone}%
\theoremstyle{thmstyletwo}%

\theoremstyle{thmstylethree}%

\begin{document}

\title[Investigations of effect of temperature and strain dependent material properties on thermoelastic damping - A generalized 3-D finite element formulation]{Investigations of effect of temperature and strain dependent material properties on thermoelastic damping - A generalized 3-D finite element formulation}


\author*[1]{\fnm{Saurabh} \sur{Dixit}}\email{saurabh\_dixit@uml.edu}

\affil*[1]{\orgdiv{Mechanical Engineering}, \orgname{University of Massachusetts Lowell}, \orgaddress{\street{1 University Ave}, \city{Lowell}, \postcode{01854}, \state{MA}, \country{USA}}}


\abstract{A comprehensive 3-D finite element formulation for the coupled thermoelastic system is proposed based on the Total Lagrangian framework to study the thermoelastic damping (TED) in small scale structures. The proposed formulation takes into account geometric nonlinearity because of large deformation and material nonlinearity where material parameters are functions of temperature and strain field.  Using the proposed finite element formulation, the TED quality factor is obtained for 1-D rod undergoing longitudinal vibrations using the eigenvalue analysis. We first validate the accuracy of the finite element implementation with previously known theoretical and numerical results. Subsequently we demonstrate the utility of the proposed numerical framework to study the effect of geometric nonlinearity, temperature and strain dependent material nonlinearity on the thermoelastic damping.In addition, the effect of internal/ external heating and different thermal boundary conditions on TED is discussed}




\maketitle

\section{Introduction}
Thermoelasticity is a field that studies the effect of coupling between the strain and temperature field. In the past,lot of work has been done in the field of thermoelasticity \cite{thomson1878,biot1956,pitarresi2003}. Since its development, it has found much significance in different areas of practical applications. For example, in the case of thermal actuators, the resonator is put into vibration with the help of external heat source \cite{huang1999,yan2004}. In addition to thermal actuators, studies on electrostatically actuated resonators show that thermoelastic coupling plays a significant role in the overall dissipative behavior of MEMS/NEMS devices \cite{zener1938,lifshitz2000}. This dissipative behavior is known as thermoelastic damping (TED).

In addition to thermoelastic damping, other dissipation mechanisms such as air loss, clamping loss, and surface loss also contribute to the overall dissipative behavior of different MEMS/NEMS devices \cite{yang2002}. Among these dissipation mechanisms, thermoelastic damping is fundamental to the system, which can not be minimized by controlling the external environment, thus put an upper limit to the extent to which quality factor can be improved. 

Thermoelastic equations are coupled in terms of strain and temperature field, which makes it difficult to solve analytically. These problems have been solved analytically by considering either one way coupling \cite{prabhakar2009} where temperature field is unaffected by strain field or problems which neglects geometric and material nonlinearities which are significant in the case of MEMS/ NEMS devices \cite{nayfeh2004,sun2009,tunvir2010,sharma20112,li20122,nourmohammadi2013,fang2013,jiao2014,zuo2016}. More accurate results are obtained when these equations are solved with numerical techniques such as finite element methods \cite{yi2007,tang2008,serra2009,basak2011,rezazadeh2015,hossain2016}. 

The effect of geometric nonlinearities on thermoelastic behavior of flexural beams considering midplane stretching in the analysis is studied \cite{zamanian2010,vahdat2011,haddadzadeh2014}. 
The effect of material nonlinearity such as temperature dependent material properties on the thermoelastic properties of nanostructures are studied in the past. Ezzat $ et$ $al. $ \cite{Ezzat2004} have discussed the role of temperature-dependent elastic modulus on thermoelastic properties of a plate using L-S theory with one thermal relaxation time. Othman $ et$ $al. $ \cite{Othman2011} have solved two-dimensional thermoelastic plate with the internal heat source and temperature-dependent elastic modulus. Abbas $ et$ $al. $ \cite{abbas2012} discussed that the temperature and stress distribution is a thermal shock problem of generalized thermoelastic half-space by using thermoelastic theory with one relaxation time. Zenkour $ et$ $al. $ \cite{Zenkour2015} have used non-local beam theory to study the effect of temperature-dependent thermal conductivity on thermoelastic properties of flexural beams. In a similar analysis, Zenkour $ et$ $al. $ \cite{Zenkour20152} has used non-local beam theory to study the effect of temperature-dependent parameters on a beam under sinusoidal thermal loading. Ren $et.al.$ \cite{ma16196390}, have looked into change in buckling behaviour of buckling due to non-uniform temperature distribution. Kim $ et$ $al. $ \cite{kim2008} have analyzed the dependence of quality factors arising due to different loss mechanisms, including TED on temperature both theoretically and analytically. Kumar $ et$ $al. $ \cite{kumar2013} have conducted experiments to study the effect of strain-dependent thermal conductivity on the quality factor of NEMS resonators. They concluded that strain could be used as a control parameter to modify the resonant frequency and the quality factor of the beam. Vujadinovi? $et.al.$ \cite{Hiller2023}, has studied the variation of quality factor across temperature in open-Loop MEMS gyroscopes. The quality factor among others has thermoelastic damping as a part of it.

During past two decades, significant amount of work has been done to study the mechanical behavior of thermoelastic bars/rods. Initially, it was shown that 1D thermoelastic solutions are stable under all physical boundary conditions \cite{kim1992}.  Gandhe $et.al.$ \cite{gandhe2015} have derived the temperature distribution in a rod fixed at both ends by considering one way coupling of heat and equilibrium equation. Semperlotti $et.al.$ \cite{Semperlotti2014} have discussed motion included heat flux due to thermoelastic waves. Bertarelli $et.al.$ \cite{bertarelli2008} have analytical studied the dynamic response of a rod rapidly heated at a point by considering one way coupling and in the end compared their results with experiments performed at CERN. Srivastava $et.al.$ \cite{srivastava2005} have compared experimental and analytical results for the relationship between volume expansion and temperature change. Jiao $et$ $al.$ \cite{jiao2014} investigated thermoelastic damping using thermal-energy method. Similar studies have discussed the thermoelastic damping in a longitudinally vibrating rod/ bars \cite{li2016,maroofi2015}.

There have been several attempts to solve the TED problems numerically. Zhang $et$ $al.$ studied and predicted thermoelastic damping in the longitudinal vibration using FEMLAB software \cite{zhang2004}. Prabhakar $et$ $al.$ coupled two dimensional heat equation with Euler-Bernoulli beam theory to obtain the quality factor in terms of infinite series solution \cite{prabhakar2008}. Guo $et$ $al.$ studied the effect of geometry of microbeam resonators on the TED using customized finite element method \cite{guo2013}.  Serra and Bonaldi \cite{serra2009} developed a finite element formulation based on 2-D Reissner?Mindlin plate theory and 3-D elastic structure. 
Parayil $et$ $al.$ presented generalized Timoshenko beam theory based finite element model to study the TED in beams with mid-plane stretching nonlinearity \cite{parayil2018}. Dixit $et.al.$ \cite{DIXIT2021106159} has studied effect of non-linearity and surface effects of size dependent thermal properties on thermoelastic damping by using non-linear beam theory. Zhang $et.al.$\cite{ZHANG20191031}, has studied thermoelastic damping in thin beam with temperature dependent material properties using perturbation methods. Zhnag $ et.al.$ \cite{ZHANG2021121576} has studied effect of large temperature on thermoelastic damping in beams using numerical methods.  Liu $et.al.$\cite{LIU2023252}, has studied one way coupled themomechanical problems under string aerothermaodynamical environment by using dynamic nonlinear thermomechanical coupling finite element algorithm. Resmi $et.al.$ \cite{resmi2021}, has studied change in quality factor of thermoelastic beam with temperature due to temperature dependent material properties in vibrating circular plates. Schiwietz $et.al.$\cite{Schiwietz2023}, has studied the change in quality factor with temperature of thermoelastic beam due to temperature dependent thermal properties by using model superposition method and compared them with experimental results. 
The similar studies can be performed in the commercial software COMSOL Multiphysics, but to get the solution one would require to provide the frequency at the beginning and the accuracy of the solution depends upon the manual scaling settings \cite{Comsol}. 

Above mentioned studies use isolated numerical techniques to solve various cases of nonlinear thermoelastic problems separately. As per the current literature, a trivial and self-contained solving technique that is capable of including various nonlinearities and effects of external/ internal heating on the thermoelastic behavior of a system is required. Zhang $et$ $al.$ \cite{zhang2019} studied experimentally the effect of temperature dependent material properties on TED and investigated that the temperature dependent material properties can be a dominant source of energy dissipation and thermal noise. To mathematically modeled this type of physical behaviour, in this paper, a comprehensive 3-D finite element formulation for thermoelasticity has been derived (see Section \ref{FEMRnD}) which is an extension to large deformation Lagrangian finite element formulation \cite{bathe1975,reddy2004}. In Section \ref{FEM}, a finite element formulation with geometric and material nonlinearity has been derived for the one-dimensional case of a longitudinally vibrating bar/rod. The present formulation has been verified with previously known results. Finally we use our finite element formulation to explore cases which have not been studied earlier, in particular we investigate the effect of geometric and material nonlinearity on the thermoelastic behavior of a vibrating rectangular bar. We summarize our findings in Section \ref{conclusion}.

\section{Finite Strain Thermoelasticity}
In continuum mechanics, when rotation, strains, and temperature gradient in a body under motion are relatively large, the assumption of classical infinitesimal strain theory ceases to work. In such cases, new definitions for stresses, strain, entropy are required. In this section, we have discussed the definitions and relationships among the quantities needed to formulate finite strain thermoelastic problems.
\subsection{Basic definitions}
\begin{figure}[h]
	\begin{center}
		\includegraphics[width=10cm]{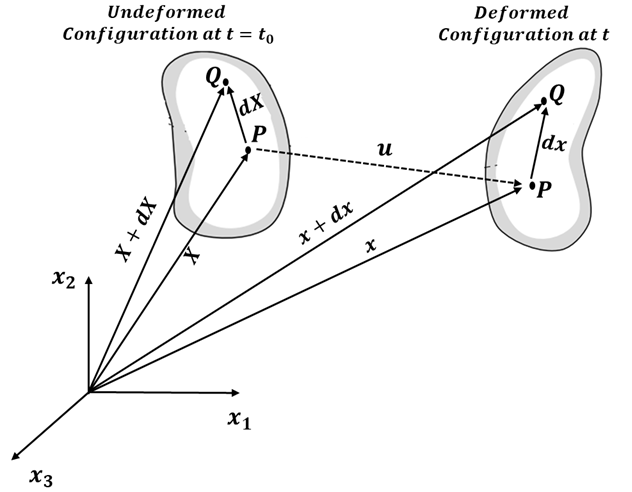}
		\caption{Motion of Continuum Body under Deformation}
		\label{fig_fs}
	\end{center}
\end{figure}
Figure \ref{fig_fs} shows a body undergoing deformation from configuration at time $t_0$ (undeformed configuration) to a configuration at time $t$ (deformed configuration) known as current configuration. In the present context, coordinates in the undeformed configuration are denoted by vector field $\bm{X}$ and in the current configuration is denoted by $\bm{x}$. Current values of physical quantities that are measured with respect to the undeformed configuration are represented with left subscript $0$ unless stated otherwise. Analytical description of the body follow two approaches. The first approach is known as the Lagrangian approach, where current coordinates $\bm{x}$ and other variables are expressed in terms of reference coordinates $\bm{X}$. The second approach is known as spatial or Eulerian mechanics, where focuses are on spatial or current coordinates. The current study uses Lagrangian description to formulate the problem.
Line element d$x$ in current configuration and d$X$ in reference configuration are related to each other by deformation gradient $\bm{F}$ $i.e.$
\begin{align}
	\mathrm{d}\bm{x}=\bm{F}.\mathrm{d}\bm{X}
\end{align}
where 
\begin{align}
	\label{dg}
	\bm{F}& =(\frac{\partial \bm{x}}{\partial \bm{X}})^T \\ \nonumber
	& =(\bm{\nabla}_0 \bm{u}+\bm{I})^T
\end{align}
where, $\nabla_0$ is the gradient operator.
Thermoelastic equations in continuum mechanics are formulated in terms of temperature and elastic strains \cite{chaves2013}. To formulate thermoelastic equations, Helmholtz free energy, which is the measure of useful work done by the system at constant temperature and volume (strains), is used as a potential function. Specific Helmholtz free energy (per unit mass) given by $\psi$ of a system is defined as \footnote{All the quantities in this section are written in the Lagrangian description unless stated otherwise.}:
\begin{flalign}
	\psi =e-T\eta 
\end{flalign}
where, $e$ and $\eta$ are internal energy density and entropy density, respectively.
By using the above definition of free energy and Clasius-Plank inequality, Second-Piola Kirchoff stress tensor $\bm{S}(\bm{E},T)$ and entropy density $\eta(\bm{E},T)$ can be represented in terms of free energy as \cite{chaves2013}
\begin{subequations}
	\begin{flalign}
		& \bm{S}(\bm{E},T)=\rho _0 \frac{\partial \psi}{\partial \bm{E}}\label{ferc}\\ 
		& \eta(\bm{E},T)=-\frac{\partial \psi}{\partial T} \label{fsrc} 
	\end{flalign}
\end{subequations}
Governing equations for coupled thermoelastic problems have been derived by using five basic balance equations viz. principle of mass balance, conservation of linear and angular momentum, first and second law of thermodynamics. Conservation of mass gives the relation $\rho=J^{-1}\rho _0$ between density in current configuration ($\rho$) and undeformed configuration($\rho_0$). Here $J$ is the jacobian. The conservation of angular momentum ensures the symmetry of stress tensor as $\bm{S}=\bm{S}^T$. Conservation of linear momentum gives the following governing equation for thermoelastic solid
\begin{flalign}
	\label{clmc}
	\bm{\nabla _0.(SF^T)}+\rho _0 \bm{b}=\rho _0 \bm{a}
\end{flalign}
where $\bm{b_0}$ is the body force per mass.
Similarly by using first and second law of thermodynamics, following governing equation can be derived
\begin{flalign}
	\label{flte}
	\rho _0T\dot{\eta}=-\bm{\nabla _0}.\bm{q_0}+\rho _0r
\end{flalign}
where, $\dot{\eta}$ is the rate of change of entropy density and $r$ is the internal heat generation per unit volume.
\subsection{Linear thermoelasticity and constitutive relations}
In linear thermoelasticity state functions such as stress tensor and entropy are linear functions of strain and temperature field. Thus according to Eq.\eqref{ferc} and Eq.\eqref{fsrc} free energy $\psi$ will be a quadratic function of strain and temperature field. Keeping this in mind free energy expanded by using Taylor series around a configuration $C_1$ characterized by strain tensor $\bm{E_1}$ and temperature field $T_1$. 
\begin{align}
	\label{fe}
	\psi(\bm{E},T)=\psi(&\bm{E_1},T_1)+\frac{\partial \psi (\bm{E_1},T_1)}{\partial \bm{E}}:(\bm{E}-\bm{E_1})+\frac{\partial \psi (\bm{E_1},T_1)}{\partial T}(T-T_1)\nonumber \\
	& +\frac{1}{2}(\bm{E}-\bm{E_1}):\frac{\partial ^2 \psi (\bm{E_1},T_1)}{\partial \bm{E} ^2}:(\bm{E}-\bm{E_1})\\
	&+\frac{1}{2}(T-T_1)\frac{\partial ^2 \psi (\bm{E_1},T_1)}{\partial \bm{E} \partial T}:(\bm{E}-\bm{E_1})+\frac{1}{2}\frac{\partial ^2 \psi (\bm{E_1},T_1)}{\partial T^2}(T-T_1)^2 \nonumber
\end{align}
By using above expression for $\psi$ and relations given in Eq.\eqref{ferc} and Eq.\eqref{fsrc} following expressions for specific entropy and stress tensor are derived:
\begin{subequations}
	\begin{flalign}
		\eta (\bm{E},T)=&-\frac{\partial \psi (\bm{E_1},T_1)}{\partial T}-\frac{\partial ^2 \psi (\bm{E_1},T_1)}{\partial T\partial \bm{E} }:(\bm{E}-\bm{E_1}) \nonumber\\
		&-(T-T_1)\frac{\partial ^2\psi (\bm{E_1},T_1)}{\partial T^2}\label{ese:a}\\
		\bm{S}(\bm{E},T)=&\rho _0\frac{\partial \psi (\bm{E_1},T_1)}{\partial \bm{E}}+\rho _0 \frac{\partial ^2 \psi (\bm{E_1},T_1)}{\partial  \bm{E}^2}:(\bm{E}-\bm{E_1}) \nonumber \\
		&+\rho _0(T-T_1)\frac{\partial ^2\psi (\bm{E_1},T_1)}{\partial \bm{E}\partial T}\label{ese:b}
	\end{flalign}
\end{subequations}

Constant terms in Taylor series (\eqref{ese:a}) at linearization point ($\bm{E_1},T_1)$ are defined as material properties at ($\bm{E_1},T_1 )$ \cite{chaves2013,clayton2013}. 
\begin{flalign}
	\label{etc}
	& \textrm{(\textbf{i}) }\bm{\prescript{1}{0}{C}}=\frac{\partial \bm{S} (\bm{E_1},T_1)}{\partial \bm{E}}\bigg|_{\dot{T}=0}=\prescript{0}{}{\rho}\frac{\partial ^2\psi (\bm{E_1},T_1)}{\partial \bm{E}^2}\\ \nonumber
	& \textrm{(\textbf{ii}) }\bm{\prescript{1}{0}{M}}=\frac{\partial \bm{S}(\bm{E_1},T_1)}{\partial T}\bigg|_{\dot{\bm{E}}=0}=\prescript{0}{}{\rho}\frac{\partial ^2\psi (\bm{E_1},T_1)}{\partial T \partial \bm{E}}\\ \nonumber
	& \textrm{(\textbf{iii}) }\prescript{1}{}{c_v}=T_1\frac{\partial \eta (\bm{E_1},T_1)}{\partial T}\bigg|_{\dot{\bm{E}}=0}=-T_1\frac{\partial ^2\psi (\bm{E_1},T_1)}{\partial T^2},
\end{flalign}
where $\bm{\prescript{1}{0}{C}}$ is a fourth order tensor called material elasticity tensor at constant temperature, $\bm{\prescript{1}{0}{M}}$ is a second order thermal stress tensor at constant strain and $\prescript{1}{}{c_v}$ is specific heat per unit mass at constant volume. Left superscript $1$ denotes that these constants are measured at $(\bm{E_1},T_1)$ while left subscript denotes that these constants relates function defined with respect to undeformed configuration $C_0$. Keeping above definitions in mind, constitutive relations for stress tensor and specific entropy can be written as
\begin{subequations}
	\label{sten}
	\begin{flalign}
		& \bm{S}(\bm{E},T)=\bm{S}(\bm{E_1},T_1)+\bm{\prescript{1}{0}{C}}:(\bm{E}-\bm{E_1})+\bm{\prescript{1}{0}{M}}(T-T_1)
		\label{sten:a}\\ 
		& \eta(\bm{E},T)=\eta (\bm{E_1},T_1)-\bm{\prescript{1}{0}{M}}(T-T_1):(\bm{E}-\bm{E_1})+\frac{\prescript{1}{}{c_v}}{T_1}(T-T_1) 
		\label{sten:b}
	\end{flalign} 
\end{subequations}
Thermal stress tensor($\bm{M}$) is usually written in terms of thermal expansion coefficients defined as $\bm{\alpha _0}=-\frac{\partial \bm{E}}{\partial T}{\bigg|}_{\bm{S}}$ which in combination with eq.\eqref{sten:b} gives
\begin{align}
	\label{tec}
	\bm{M}=-\bm{C}:\bm{\alpha _0}
\end{align}

As a summary following governing equations with corresponding constitutive relations completes the present section (for reference see eqs.\eqref{clmc}, \eqref{flte} and \eqref{sten}).
\begin{subequations}
	\label{ftee}
	\begin{flalign}
		& \mathrm{div}(\bm{S}.\bm{F}^T)+\rho _0\bm{b}=\rho _0\bm{\ddot{u}} \label{ftee:a}\\
		\mathrm{where \,\,\,\,}& \bm{S}(\bm{E},T)=\bm{S}(\bm{E_1},T_1)+\prescript{1}{0}{\bm{C}}:(\bm{E}-\bm{E_1})-\prescript{1}{0}{\bm{C}}:\prescript{1}{0}{\bm{\alpha}}(T-T_1)
		\label{ftee:c}\\
		& \rho _0T\dot{\eta}=-\bm{\nabla _0}.\bm{q_0}+\rho _0 r \label{ftee:b}\\
		\mathrm{where \,\,\,\,}
		& \dot{\eta}(\bm{E},T)=\frac{1}{\rho _0}\prescript{1}{0}{\bm{C}}:\prescript{1}{0}{\bm{\alpha}}:\dot{\bm{E}}+\frac{\prescript{1}{}c_v}{T_1}\dot{T} \label{ftee:d}    \end{flalign}
\end{subequations}

\section{Finite Element Formulation of Finite Strain Linear Thermoelasticity}
\label{FEMRnD}
Given the complexity of coupled thermoelastic equations as derived in Eq.\eqref{ftee}, the task of finding an analytical solution is a trivial one. The finite element method has extensively been used to solve small deformation thermoelastic problems in the past. Among others, co-rotational and Total/ Updated Lagrangian finite element techniques have extensively been used to solve large deformation problems \cite{bathe1975, crisfield1990}. In this section, Total Lagrangian finite element formulation for thermoelastic problems, similar to the one described in previous works on large deformation\cite{bathe1975, reddy2004, bathe2006}, has been presented.

\subsection{Basic definitions}
Similar to the Updated/Total Lagrangian formulation \cite{bathe1975}, we propose that thermoelastic problem problems can be solved by dividing the heat input into smaller steps in addition to smaller load steps. Each step should be such that the current state can be linearized around the previous step. In the Total Lagrangian formulation, the stress, strain and heat flux are expressed with respect to the initial configuration.

\begin{figure}[h]
	\begin{center}
		\includegraphics[width=12cm]{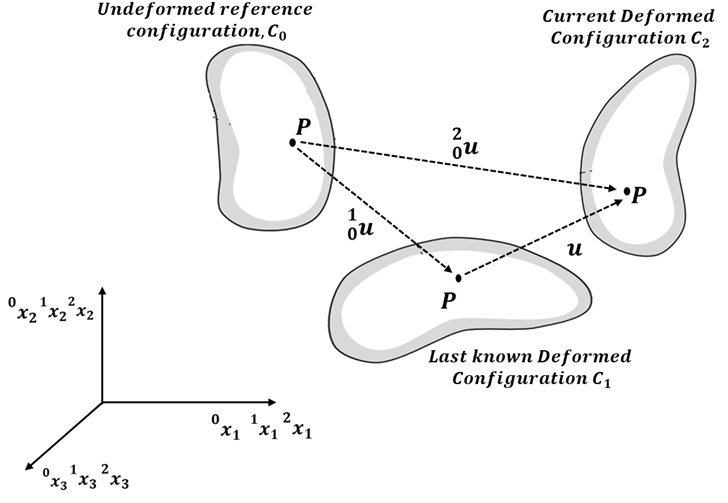}
		\caption{Different Configurations for Lagrangian Formulation}
		\label{fig_UL}
	\end{center}
\end{figure}

Figure \ref{fig_UL} shows a body undergoing deformation through different configurations. Here configuration $C_0$ represents the undeformed reference configuration, $C_2$ represents the current deformed configuration and $C_1$ represents last known deformed configuration. It is assumed that all the configuration up to $C_1$ is known while deformation and temperature change from configuration $C_1$ to $C_2$ is small, although it can be arbitrary large till configuration $C_1$. Following notation conventions shall be used throughout the finite element formulation \cite{bathe1975,reddy2004}.\\
(i) Left superscript is used to indicate the configuration in which it occurs.\\
(ii) A left subscript is used to indicate the configuration with respect to which it is measured.\\
(iii) Left superscript shall be omitted for incremental quantities occurring between $C_1$ and $C_2$.\\
For example, quantity $^i _j Q$ implies that the $Q$ is occurring in configuration $i$ and is measured with respect to configuration $j$, and $_j Q$ implies that the incremental quantity $Q$ which is measured with respect to configuration $j$.

Since constitutive relations for stress tensor and entropy as derived in Eq.\eqref{ftee} are linear with respect to strain and temperature field, the current state of stress and entropy can be written as a summation of stresses and entropy at a given state $C_1$ and additional terms. In Total/ Updated Lagrangian formulation, these additional terms are known as incremental quantities $i.e., $
\begin{subequations}
	\begin{flalign}
		& \prescript{2}{0}{S_{ij}} =\prescript{1}{0}{S_{ij}} +\prescript{}{0}{S_{ij}}\label{isstr:a}\\
		& \prescript{2}{0}{\eta}=\prescript{1}{0} {\eta}+\prescript{}{0}{\eta} \label{isstr:b}\\
		& \prescript{2}{0}{Q_{i}} =\prescript{1}{0} Q_{i}+\prescript{}{0} Q_{i} \label{isstr:c}
	\end{flalign}
\end{subequations} 
where $\prescript{}{0}S_{ij}$, $\prescript{}{0}{\eta}$ and $\prescript{}{0}{Q_i}$ are called as incremental stress, incremental entropy and incremental heat flux respectively. Following Eq.\eqref{ftee}, incremental stress/ entropy can be written in terms of incremental strain, $\prescript{}{0}{\epsilon}$ and incremental temperature $\theta$ as:
\begin{subequations}
	\begin{flalign}
		& \prescript{}{0}S_{ij} =\prescript{1}{0}C_{ijkl}\prescript{}{0}\epsilon _{kl}-\prescript{1}{0}M_{ij}\theta \label{isr1}\\
		& \prescript{}{0}{\eta}=-\frac{1}{\prescript{0}{}{\rho}}\prescript{1}{0}M_{ij}\prescript{}{0}\epsilon _{kl}+\frac{\prescript{1}{}{c_v}}{\prescript{1}{}T}\theta \label{ienr1}\\
		& \prescript{}{0}{q_i}=-\prescript{1}{0}K_{ij}\partial _j \theta \label{ihfr1}
	\end{flalign}
\end{subequations}
Coordinates of a point in undeformed configuration $C_0$, last known configuration $C_1$ and current configuration $C_2$ are denoted by $\bm{\prescript{0}{}{x}}$, $\bm{\prescript{1}{}{x}}$ and $\bm{\prescript{2}{}{x}}$, respectively. Displacement of a point in different configuration are represented as
\begin{flalign}
	\bm{\prescript{2}{0}{u}}=\bm{\prescript{2}{}{x}}-\bm{\prescript{0}{}{x}} \nonumber \\
	\bm{\prescript{1}{0}{u}}=\bm{\prescript{1}{}{x}}-\bm{\prescript{0}{}{x}} 
\end{flalign}
Relationship between incremental displacement $\bm{u}=\bm{\prescript{2}{0}{u_2}}-\bm{\prescript{2}{0}{u_1}}$ and incremental strain is given by \cite{bathe1975,reddy2004}
\begin{flalign}
	\prescript{2}{0}{E_{ij}}& =\frac{1}{2}\left(\frac{\partial \prescript{2}{0}{u_i}}{\partial \prescript{0}{}{x_j}}+\frac{\partial \prescript{2}{0}{u_j}}{\partial \prescript{0}{}{x_i}}+\frac{\partial \prescript{2}{0}{u_k}}{\partial \prescript{0}{}{x_i}}\frac{\partial \prescript{2}{0}{u_k}}{\partial \prescript{0}{}{x_j}}\right) \nonumber \\
	& =\frac{1}{2}\left[\frac{\partial (\prescript{1}{0}{u_i}+u_i)}{\partial \prescript{0}{}{x_j}}+\frac{\partial (\prescript{1}{0}{u_j}+u_j)}{\partial \prescript{0}{}{x_i}}+\frac{\partial (\prescript{1}{0}{u_k}+u_k)}{\partial \prescript{0}{}{x_i}}\frac{\partial (\prescript{1}{0}{u_k}+u_k)}{\partial \prescript{0}{}{x_j}}\right] \nonumber \\
	&=\frac{1}{2}\left(\frac{\partial \prescript{1}{0}{u_i}}{\partial \prescript{0}{}{x_j}}+\frac{\partial \prescript{1}{0}{u_j}}{\partial \prescript{0}{}{x_i}}+\frac{\partial \prescript{1}{0}{u_k}}{\partial \prescript{0}{}{x_i}}\frac{\partial \prescript{1}{0}{u_k}}{\partial \prescript{0}{}{x_j}}\right) \nonumber \\
	&\,\,\,\,\,\,+\frac{1}{2}\left(\frac{\partial u_i}{\partial \prescript{0}{}{x_j}}+\frac{\partial \prescript{1}{0}{u_j}}{\partial \prescript{0}{}{x_i}}+\frac{\partial \prescript{1}{0}{u_k}}{\partial \prescript{0}{}{x_i}}\frac{\partial u_k}{\partial \prescript{0}{}{x_j}}+\frac{\partial u_k}{\partial \prescript{0}{}{x_i}}\frac{\partial \prescript{1}{0}{u_k}}{\partial \prescript{0}{}{x_j}}\right)+\frac{1}{2}\left(\frac{\partial u_k}{\partial \prescript{0}{}{x_i}}\frac{\partial u_k}{\partial \prescript{0}{}{x_j}}\right)\nonumber \\
	&=\prescript{1}{0}{E_{ij}}+\prescript{}{0}{\epsilon _{ij}}
\end{flalign}
For simplification incremental strain $\prescript{}{0}{\epsilon _{ij}}$ can be written as a summation of its linear and nonlinear part (linearity and nonlinearity in terms of unknown displacement $u$) as
\begin{flalign}
	\prescript{}{0}{\epsilon _{ij}}=\prescript{}{0}{e_{ij}}+\prescript{}{0}{\gamma _{ij}} \label{tlln}
\end{flalign}
where 
\begin{subequations}
	\label{tlis}
	\begin{flalign}
		& \prescript{}{0}{e_{ij}}=\frac{1}{2}\left(\frac{\partial u_i}{\partial \prescript{0}{}{x_j}}+\frac{\partial \prescript{1}{0}{u_j}}{\partial \prescript{0}{}{x_i}}+\frac{\partial \prescript{1}{0}{u_k}}{\partial \prescript{0}{}{x_i}}\frac{\partial u_k}{\partial \prescript{0}{}{x_j}}+\frac{\partial u_k}{\partial \prescript{0}{}{x_i}}\frac{\partial \prescript{1}{0}{u_k}}{\partial \prescript{0}{}{x_j}}\right) \label{tllis}\\
		& \prescript{}{0}{\gamma _{ij}}=\frac{1}{2}\left(\frac{\partial u_k}{\partial \prescript{0}{}{x_i}}\frac{\partial u_k}{\partial \prescript{0}{}{x_j}}\right) \label{tlnlis}
	\end{flalign}
\end{subequations}
Note that above definitions of incremental stress/strain and incremental temperature are equivalent to the situation where configuration $C_1$ is assumed as linearizion point in eq\eqref{ftee}. Since the changes from $C_1$ to $C_2$ is very small it perfectly fits with the Taylor series assumptions of neglecting higher order terms in eq.\eqref{fe}. While classic thermoelastic formulation is based upon the assumption that temperature variation from undeformed state is very small so it can be linearized around its undeformed state, present way of formulation allows for relatively large temperature variation and deformations. One main benefit of the present formulation is that material constants $\alpha _{ij}$, $C_{ijkl}$, $K_{ij}$ and $c_v$ are functions of temperature and strain field at known confguration $C_1$ and thus it allows easy inclusion of such material non-linearities.
\subsection{Formulation } 
In this section Total Lagrangian formulation of  coupled thermoelastic equations (see Eq. \eqref{ftee}) has been carried out. A principle of virtual work method is used to derive the equations of the Lagrangian incremental description of motion. In the case of Eq. \eqref{ftee:a} a virtual displacement ($\delta \bm{u}$) is considered to obtain the incremental equations.
\begin{flalign}
	\int_{\prescript{0}{}{V}}((\prescript{2}{0}{S_{ij}}\prescript{2}{0}{F_{kj}})_{,k}+\prescript{0}{}{\rho}b_i)\delta u_id\prescript{0}{}{V} =\int_{\prescript{0}{}{V}}\prescript{0}{}{\rho} \ddot{u}_i\delta u_id\prescript{0}{}{V}
\end{flalign}
By using definitions of $\delta u_{i,k}$ (from Eq.\eqref{dg} ), symmetric tensor $\prescript{2}{0}{S_{ij}}$ and Green-Lagrangian strain we get,
\begin{flalign}
	\label{wfeeq}
	& \int_{\prescript{0}{}{V}}\prescript{0}{}{\rho} \ddot{u}_i\delta u_id\prescript{0}{}{V}+\int_{\prescript{0}{}{V}}\prescript{2}{0}{S_{ij}}\delta \prescript{2}{0}{E_{ij}}d\prescript{0}{}{V}-\delta \prescript{2}{0}{R}=0 \\
	&\mathrm{where \,\,\,\,}\delta \prescript{2}{0}{R}=\int_{\prescript{0}{}{V}}\prescript{0}{}{\rho}b_i\delta u_id\prescript{0}{}{V}+\int_{\prescript{0}{}{A}}\prescript{2}{0}{t_i}\delta u_id\prescript{0}{}{A}=0\nonumber
\end{flalign}

and $\prescript{2}{0}{t_i}$ is external surface traction per unit undeformed area. \\
Since $\prescript{1}{0}{E_{ij}}$ is already known, $\delta \prescript{2}{0}{E_{ij}}=\delta \prescript{}{0}{\epsilon _{ij}}$. By using definitions of incremental strains and stresses from Eq.\eqref{isstr:a} and Eq.\eqref{tlln} and constitutive relation from Eq.\eqref{isr1}
\begin{flalign}
	\label{tleeqea}
	\delta \prescript{2}{0}{R}&=\int _{\prescript{0}{}{V}}\prescript{0}{}{\rho}\ddot{u}_i \delta u_i d\prescript{0}{}{V}+\int _{\prescript{0}{}{V}}(\prescript{1}{0}{S_{ij}}+\prescript{}{0}{S_{ij}})\delta (\prescript{}{0}{\epsilon _{ij}}) d\prescript{0}{}{V} \nonumber \\
	&=\int _{\prescript{0}{}{V}}\prescript{0}{}{\rho}\ddot{u}_i \delta u_i d\prescript{0}{}{V}+\int _{\prescript{0}{}{V}}\prescript{1}{0}{S_{ij}}\delta (\prescript{}{0}{\epsilon _{ij}}) d\prescript{0}{}{V}+\prescript{}{0}{S_{ij}}\delta (\prescript{}{0}{\epsilon _{ij}}) d\prescript{0}{}{V} \nonumber \\
	&\,\,\,\,\,\,\,\, +\left[\prescript{1}{0}{C_{ijkl}} \prescript{}{0}{\epsilon _{kl}}+ \prescript{1}{0}{M_{ij}}\theta\right]\delta (\prescript{}{0}{\epsilon _{ij}}) d\prescript{0}{}{V}
\end{flalign}
Note that till now while deriving above equation no assumption has been made in the preview of linear thermo elasticity. To make things computationally solvable we make use of our initial assumption that strains from $C_1$ to $C_2$ are very small. In this case
\begin{flalign}
	\prescript{1}{0}{C_{ijkl}} \prescript{}{0}{\epsilon _{kl}}=\prescript{1}{0}{C_{ijkl}} \prescript{}{0}{e _{kl}}, \, \,\,\,\, \delta (\prescript{}{0}{\epsilon _{ij}})=\delta (\prescript{}{0}{e _{ij}}).
\end{flalign}
By making use of above assumption, Eq.\eqref{tleeqea} can be rewritten as
\begin{flalign}
	\label{tleewf}
	& \int _{\prescript{0}{}{V}}\prescript{0}{}{\rho}\ddot{u}_i \delta u_i d\prescript{0}{}{V}+ \int _{\prescript{0}{}{V}}\prescript{1}{0}{S_{ij}}\delta (\prescript{}{0}{\gamma _{ij}})d\prescript{0}{}{V}+\int _{\prescript{0}{}{V}}\prescript{1}{0}{C_{ijkl}} \prescript{}{0}{e _{kl}}\delta (\prescript{}{0}{e _{ij}}) d\prescript{0}{}{V} \nonumber \\
	&\,\,\,\,\, + \int _{\prescript{0}{}{V}}\prescript{1}{0}{M_{ij}}\theta \delta (\prescript{}{0}{e _{ij}}) d\prescript{0}{}{V} =\delta \prescript{2}{0}{R}-\int _{\prescript{0}{}{V}}\prescript{1}{0}{S_{ij}}\delta (\prescript{}{0}{e _{ij}})d\prescript{0}{}{V}
\end{flalign}
Similarly, Lagrangian incremental description of Eq.\eqref{ftee:b} is derived by using $\delta \theta$.
\begin{flalign}
	\int_{\prescript{0}{}{V}} \left[\prescript{0}{}{\rho}\prescript{2}{}{T}\prescript{2}{0}{\dot{\eta}}+\partial _i\prescript{2}{0}{q_i}-\prescript{0}{}{\rho} r\right]\delta \theta d\prescript{0}{}{V}=0 
\end{flalign}
Following similar procedure which is used to derive incremental equation of equilibrium equation (Eq.\eqref{wfeeq}) we get,
\begin{flalign}
	\label{wfheq}
	& \int_{\prescript{0}{}{V}} \prescript{0}{}{\rho}\prescript{2}{}{T}\prescript{2}{0}{\dot{\eta}}\delta \theta d\prescript{0}{}{V}-\int_{\prescript{0}{}{V}}\prescript{2}{0}{q_i}\delta \theta _{,i}d\prescript{0}{}{V}-\delta \prescript{2}{0} {H}=0 \\
	&\mathrm{where \,\,\,\,}\delta\prescript{2}{0}{H}=\int_{\prescript{0}{}{V}}\prescript{0}{}{\rho} r\delta \theta d\prescript{0}{}{V}-\int_{\prescript{0}{}{A}}\prescript{2}{0}{q_i}\prescript{0}{}{n_i}\delta \theta d\prescript{0}{}{A} \nonumber 
\end{flalign}
This equation is derived without any assumptions and since temperature change is very small from configuration $C_1$ to $C_2$ (Note that this assumption is valid when initial temperature is large enough to ignore the incremental temperature. For large enough initial temperature, load and heat source steps can be chosen to make this assumption valid). In which case we may assume
\begin{flalign}
	\prescript{0}{}{\rho}\prescript{2}{}{T}\prescript{2}{0}{\dot{\eta}} \approx \prescript{0}{}{\rho}\prescript{1}{}{T} \prescript{2}{0}{\dot{\eta}}
\end{flalign}
By using above approximation and definition of incremental entropy and heat flux (Eq.\eqref{isstr:b} and Eq.\eqref{isstr:c}), Eq.\eqref{wfheq} can be rewritten as
\begin{flalign}
	\label{tlheqqa}
	\int_{\prescript{0}{}{V}} \prescript{0}{}{\rho}\prescript{1}{}{T}(\prescript{1}{0}{\dot{\eta}}+\prescript{}{0}{\dot{\eta}})\delta \theta d\prescript{0}{}{V}-\int_{\prescript{0}{}{V}}(\prescript{1}{0}{q_i}+\prescript{}{0}{q_i})\delta \theta _{,i} d\prescript{0}{}{V}-\delta (\prescript{2}{0}{H})=0
\end{flalign}
Next we replace incremental entropy and heat flux from constitutive relations as derived in Eq.\eqref{ienr1} and Eq.\eqref{ihfr1}, 
\begin{flalign}
	\delta (\prescript{2}{0}{H})
	&=\int_{\prescript{0}{}{V}}\prescript{0}{}{\rho}c_v\dot{\theta} \delta \theta d\prescript{0}{}{V}-\int_{\prescript{0}{}{V}}\prescript{1}{}{T}\prescript{1}{0}{M_{ij}}\prescript{}{0}{\dot{\epsilon} _{ij}}\delta \theta d\prescript{0}{}{V}+\int_{\prescript{0}{}{V}}\prescript{1}{0}{K_{ij}}\theta _{,j}\delta \theta _{,i} d\prescript{0}{}{V} \nonumber \\
	&\,\,\,\,+\int_{\prescript{0}{}{V}}\prescript{0}{}{\rho}\prescript{1}{}{T}\prescript{1}{0}{\dot{\eta}}-\int_{\prescript{0}{}{V}}\prescript{1}{0}{q_i}\delta \theta _{,i} d\prescript{0}{}{V}
\end{flalign}
By using initial assumption, incremental displacements are comparatively small ($\prescript{1}{0}{M_{ij}}\prescript{}{0}{\dot{\epsilon} _{ij}}\approx \prescript{1}{0}{M_{ij}}\prescript{}{0}{\dot{e} _{ij}}$), we get 
\begin{flalign}
	\label{tlhewf}
	& \int_{\prescript{0}{}{V}}\prescript{0}{}{\rho}c_v\dot{\theta} \delta \theta d\prescript{0}{}{V}-\int_{\prescript{0}{}{V}}\prescript{1}{}{T}\prescript{1}{0}{M_{ij}}\prescript{}{0}{\dot{e} _{ij}}\delta \theta d\prescript{0}{}{V} \nonumber \\
	&\,\,\,\,\,\,+\int_{\prescript{0}{}{V}}\prescript{1}{0}{K_{ij}}\theta _{,j}\delta \theta _{,i} d\prescript{0}{}{V}=\delta (\prescript{2}{0}{H})+\int_{\prescript{0}{}{V}}\prescript{1}{0}{q_i}\delta \theta _{,i} d\prescript{0}{}{V}-\int_{\prescript{0}{}{V}}\prescript{0}{}{\rho}\prescript{1}{}{T}\prescript{1}{0}{\dot{\eta} d\prescript{0}{}{V}}
\end{flalign}
Equation \eqref{tleewf} and Eq. \eqref{tlhewf}  forms the basis for the finite element model for nonlinear two-way coupled thermoelastic problems.
\section{Finite Element Formulation of One Dimensional Case}
\label{FEM}
In this section finite element equations for one dimensional case is derived. Stresses in non-axial directions are assumed to be zero. For simplicity subscripts used to denote tensor and vector quantities are omitted ($i.e.$ $S_{ij}$ is denoted by $S$ and $u_i$ is denoted by $u$). By using Eq.\eqref{tleewf} and Eq. \eqref{tlhewf} the incremental Lagrangian equations for one dimensional case can be written as follows:
\begin{flalign}
	\label{tl1deeq}
	&\int _{\prescript{0}{}{V}}\prescript{0}{}{\rho}\ddot{u}dud\prescript{0}{}{V}+\int _{\prescript{0}{}{V}}\prescript{1}{0}{S}\delta(\prescript{}{0}{\gamma})d\prescript{0}{}{V}+\int _{\prescript{0}{}{V}}\prescript{1}{0}{C}\prescript{}{0}{e}\delta (\prescript{}{0}{e})d\prescript{0}{}{V} \nonumber \\
	&\,\,\,\, +\int _{\prescript{0}{}{V}}\prescript{1}{0}{\alpha} \prescript{1}{0}{C}\theta \delta (\prescript{}{0}{\gamma})d\prescript{0}{}{V}=\delta \prescript{2}{0}{R}-\int _{\prescript{0}{}{V}}\prescript{1}{0}{S}\delta (\prescript{}{0}{e})d\prescript{0}{}{V}
\end{flalign} 
\begin{flalign}
	&\int_{\prescript{0}{}{V}} {\prescript{0}{}{\rho}}\prescript{1}{}{c_v}\dot{\theta}\delta \theta d\prescript{0}{}{V}-\int _{\prescript{0}{}{V}}\prescript{1}{0}{\alpha}\prescript{1}{}{T}\prescript{1}{0}{C} \prescript{}{0}{\dot{e}}\delta \theta d\prescript{0}{}{V}+\int_{\prescript{0}{}{V}}\prescript{1}{0}{K}\theta '\delta \theta 'd\prescript{0}{}{V} \nonumber \\
	&\,\,\,\, =\delta (\prescript{2}{0}{H})+\int _{\prescript{0}{}{V}}\prescript{1}{0}{q}\delta \theta 'd\prescript{0}{}{V}-\int _{\prescript{0}{}{V}}\prescript{0}{}{\rho}\prescript{1}{}{T}\prescript{1}{0}{\dot{\eta}}\delta \theta d\prescript{0}{}{V}
\end{flalign}
Linear and nonlinear part of strains as defined in Eq.\eqref{tllis} and eq.\eqref{tlnlis} can be written in terms of displacement $u$ as
\begin{flalign}
	& e=\frac{\partial u}{\partial \prescript{0}{}{x}}+\frac{\partial \prescript{1}{0}{u}}{\partial \prescript{0}{}{x}}\frac{\partial u}{\partial \prescript{0}{}{x}},\,\,\,\, \gamma=\frac{1}{2}\left(\frac{\partial u}{\partial \prescript{0}{}{x}}\right) ^2\\
	& \delta e=\frac{\partial \delta u}{\partial \prescript{0}{}{x}}+\frac{\partial \prescript{1}{0}{u}}{\partial \prescript{0}{}{x}}\frac{\partial \delta u}{\partial \prescript{0}{}{x}},\,\,\,\,\,
	\delta \gamma=\frac{\partial u}{\partial \prescript{0}{}{x}}\frac{\partial \delta u}{\partial \prescript{0}{}{x}}
\end{flalign}
The semidiscrete finite element model (spatial discretization only)
is obtained after implementing the Galerkin based
finite element procedure by assuming the following approximation 	
for the dependent variable in an element
\begin{flalign}
	u=\sum\limits_{i=1}^{m}\psi _i(\prescript{0}{}{x})u_i,\,\,\,\,\theta =\sum\limits_{i=1}^{n}\phi _i(\prescript{0}{}{x})\theta _i
\end{flalign}
where, $\psi_i$ and $\phi_i$ are the Lagrangian interpolation functions. The resulting system of coupled nonlinear ordinary differential
equations for a typical element can be expressed in matrix form
as follows: 
\begin{flalign}
	\label{tl1Dmeeq}
	[M^{uu}]\{\ddot{u}\}+[K^{uu}]\{u\}+[K^{u\theta}]\{\theta \}=\{f^u\}
\end{flalign}
and 
\begin{flalign}
	\label{tl1Dmheq}
	[D^{\theta \theta}]\{\dot{\theta}\}+[D^{\theta u}]\{\dot{u}\}+[K^{\theta \theta}]\{\theta\}=\{h\}
\end{flalign}
where
\begin{flalign}
	& M^{uu}_{ij}=\int _{\prescript{0}{}{V}}\prescript{0}{}{\rho}\psi _i \psi _j d\prescript{0}{}{V},\,\,\,\,K^{uu}_{ij}=\int _{\prescript{0}{}{V}}\prescript{1}{0}{S}\psi ^{'} _i \psi ^{'} _j d\prescript{0}{}{V}+\int _{\prescript{0}{}{V}}\prescript{1}{0}{C}(1+\prescript{1}{0}{u}')^2\psi ^{'} _i \psi ^{'} _j d\prescript{0}{}{V} \nonumber \\
	& K^{u\theta}_{ij}=-\int _{\prescript{0}{}{V}}\prescript{1}{0}{\alpha} \prescript{1}{0}{C} (1+\prescript{1}{0}{u}')\psi ^{'} _i \phi _j d\prescript{0}{}{V},\nonumber \\
	&f^u_i=\int _{\prescript{0}{}{V}}\prescript{0}{}{\rho}b\psi _i d\prescript{0}{}{V}+\int _{\prescript{0}{}{A}}\prescript{2}{0}{t}\psi _id\prescript{0}{}{A}-\int _{\prescript{0}{}{V}}\prescript{1}{0}{S}(1+\prescript{1}{0}{u}')\psi ^{'} _id\prescript{0}{}{V}. \nonumber
\end{flalign}
\begin{flalign}
	&[D^{\theta \theta}]=\int_{\prescript{0}{}{V}} {\prescript{0}{}{\rho}}c_v\phi _i\phi _j d\prescript{0}{}{V},\,\,\,\,[D^{\theta u}]=-\int _{\prescript{0}{}{V}}\prescript{1}{0}{\alpha}\prescript{1}{}{T}\prescript{1}{0}{C}  (1+\prescript{1}{0}{u}')\phi _i\psi _j'd\prescript{0}{}{V} \nonumber \\
	&[K^{\theta \theta}]=\int_{\prescript{0}{}{V}}\prescript{1}{0}{K}\phi _i '\phi _j 'd\prescript{0}{}{V} \nonumber \\
	&\{h^{\theta}\}=\int _{\prescript{0}{}{V}}\prescript{0}{}{\rho}r\phi _i d\prescript{0}{}{V}-\int _{\prescript{0}{}{A}}\prescript{2}{0}{q}\phi _i d\prescript{0}{}{A}+\int _{\prescript{0}{}{V}}\prescript{1}{0}{q}\phi _i 'd\prescript{0}{}{V}-\int _{\prescript{0}{}{V}}\prescript{0}{}{\rho}\prescript{1}{}{T}\prescript{1}{0}{\dot{\eta}}\phi _i d\prescript{0}{}{V} \nonumber
\end{flalign}
Eq.\eqref{tl1Dmeeq} and Eq.\eqref{tl1Dmheq} can be combined and written as second order equations as
\begin{flalign}
	\begin{bmatrix}
		M^{uu} & [0]_{mxn}  \\
		[0]_{nxm} & [0]_{nxn}
	\end{bmatrix}
	\begin{bmatrix}
		\{\ddot{u}\} \\
		\{\ddot{\theta}\} 
	\end{bmatrix}
	+
	\begin{bmatrix}
		[0]_{mxm} & [0]_{mxn}  \\
		D^{\theta u} & D^{\theta \theta}
	\end{bmatrix}
	\begin{bmatrix}
		\{\dot{u}\} \\
		\{\dot{\theta}\}
	\end{bmatrix}
	+\begin{bmatrix}
		K^{uu} & K^{u\theta}  \\
		[0]_{nxm} & K^{\theta \theta}
	\end{bmatrix}
	\begin{bmatrix}
		\{u\} \\
		\{\theta\} 
	\end{bmatrix}
	=
	\begin{bmatrix}
		\{f^u\} \\
		\{h^{\theta}\}
	\end{bmatrix}
\end{flalign}
The elemental matrices derived are then assembled to form the global system of coupled equations. The eigenvalue analysis of these coupled equation is carried out to study the thermoelastic damping in the one dimensional rod undergoing longitudinal vibrations.
\section{Numerical Simulations and Discussions}
\label{RnD}
In this section, the applicability of the proposed formulation has been verified, and effects due to different nonlinearities have been studied for the one-dimensional case. A rectangular bar of dimension 100nm x 10nm x 10nm has been taken as a test case for the rest of the section unless stated otherwise. 
In this section, stationary and eigenvalue problem of a longitudinal vibrating rectangular bar (Si material) is studied.
The material properties of Si at 300K have been given in Table \ref{tmc}   
\begin{center}
	\begin{table}
		\centering
		\caption {Material Constants for Si at 300$^0$K} \label{tmc} 
		\begin{tabular}{c c c}
			\hline
			Vairable & Physical Description & Value\\ [0.5ex] 
			\hline\hline
			$Y_0$ & Young's modulus & 165$GPa$ \\  
			$\nu$ & Poisson's ratio & 0.22 \\
			$\rho _0$ & Density & 2300$Kg/m^3$ \\
			$\alpha _0$ & Thermal expansion coefficient & 2.6x$10^{-6}k^{-1}$\\
			$k_0$ & Thermal conductivity & 159$Wm^{-1}K^{-1}$\\
			$c_v ^0$ & Specific heat capacity & 713$JKg^{-1}K^{-1}$\\ 
			$T_0$ & Initial temperature & 300K\\ [1ex]
			\hline
		\end{tabular}
	\end{table}    
\end{center} 
\subsection{Numerical validation of current formulation}
\label{bm} 
\begin{figure}
	\centering
	\includegraphics[width=0.5\textwidth]{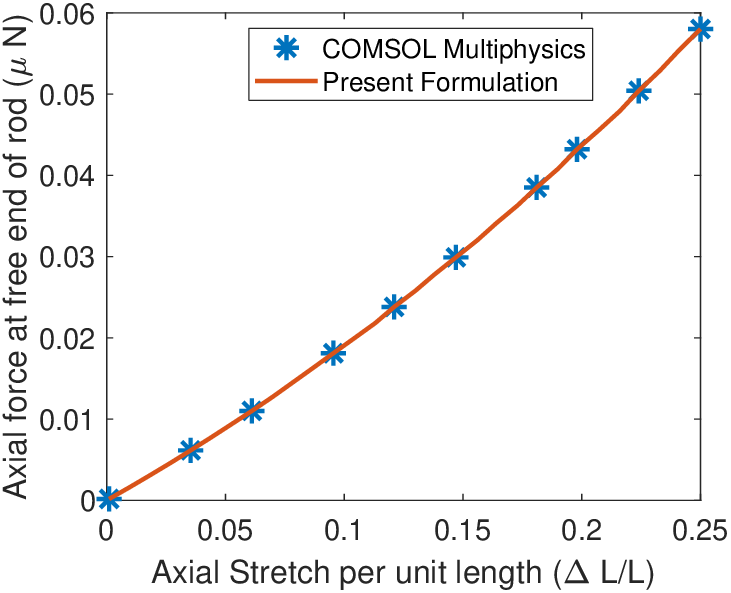}
	\caption{Comparison of force displacement characteristics for geometric nonlinear case}
	\label{fig:ReacForceVsStretch}
\end{figure}

\begin{figure}
	\centering
	\includegraphics[width=0.5\textwidth]{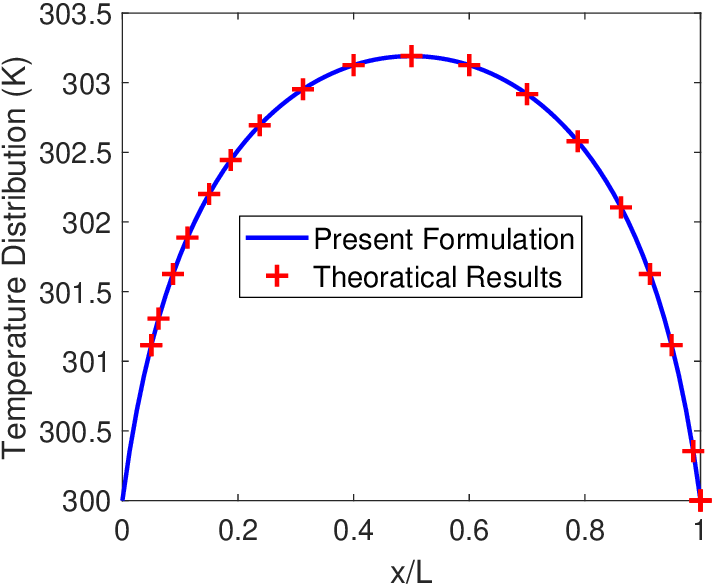}
	\caption{Comparison of temperature distribution for the system with temperature dependent thermal conductivity of the form $k=k_0e^{0.5(T-T_0)}$}
	\label{fig:TempVsLength}
\end{figure}

As discussed earlier, present formulation is capable of solving any general nonlinear thermoelastic problem where the nonlinearities may be arising from nonlinear strains or strain/ temperature dependent elastic and thermal constants. A finite element code is written in MATLAB based on the formulation given in Section \ref{FEM}. The validity of proposed formulation has been established by benchmarking the cases where only one of such nonlinearities is turned on (geometric nonlinearity or material nonlinearity) while the coupling behavior has been benchmarked for linear case only.
To validate the geometric nonlinearity part, the results obtained from the current formulation are compared with the COMSOL Multiphysics simulation results. 
A fixed-free rectangular cross-section bar with dimensions mentioned in Table \ref{tmc} undergoing large deformation in logitudinal direction is analysed using current formulation and COMSOL Multiphysics. Figure \ref{fig:ReacForceVsStretch} compares the end force versus displacement of a bar. It can be seen that the results are in perfect match with the COMSOL Multiphysics results.
\\
Similarly, to validate the material nonlinearity part, a steady-state heat conduction problem of a fixed-fixed rectangular bar with isothermal end conditions is considered. 
A temperature dependent thermal conductivity $k=k_0e^{0.5(T-T_0)}$ and the internal heat generation ($r=10^{18}$ $W/m^3$) are chosen for this simulation. A steady state temperature distribution of an exponentially varying coefficient of thermal conductivity can be anaytically determined by
\begin{flalign}
	T=T_0+\frac{1}{\beta}\mathrm{ln}\left(1+\frac{\beta rx(x-L0)}{2k_0}\right)
\end{flalign}
Figure \ref{fig:TempVsLength} shows the comparison of temperature distribution of the bar obtained using analytical calculations and FE simulations. It can be seen that the FE results are in good agreement with the analytical results.

\begin{figure}
	\centering
	\subfloat[Comparison of inverse quality factor vs length with \cite{li2016}]{
		\includegraphics[width=0.42\textwidth]{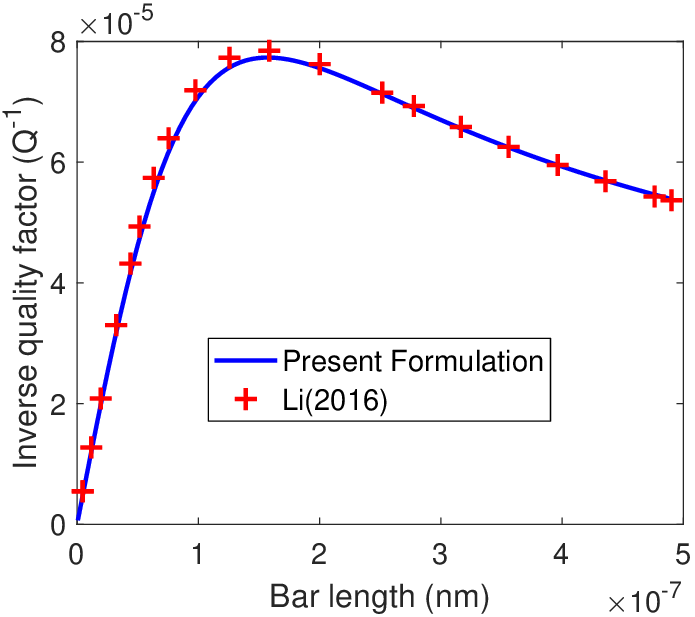}
		\label{fig:QiVsLength}}
	\quad
	\subfloat[Comparison of inverse quality factor vs length with \cite{jiao2014}]{
		\includegraphics[width=0.45\textwidth]{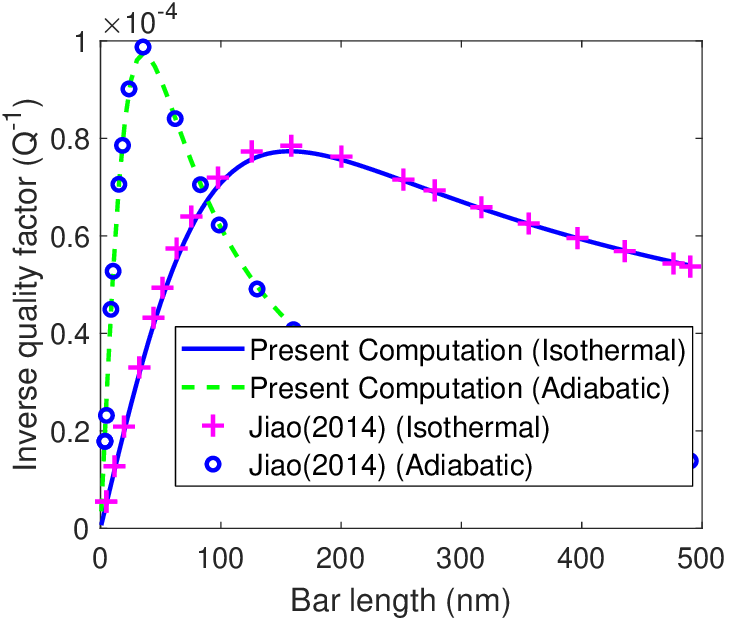}
		\label{fig:QiVsLengthJ}}
	\caption{Comparison of thermoelastic damping quality factor obtained using present formulation with Li \cite{li2016} and Jiao \cite{jiao2014} }
	\label{QiVsLengthC}
\end{figure}

The efficacy of thermoelastic coupling is checked for case without geometric and material nonlinearity. A coupled linear thermoelastic analysis of a fixed-free bar is performed using current formulation. The dissipative behavior of longitudinally vibrating bar arising from thermoelastic damping are then compared with the previously reported results by Li $ et$ $al$. \cite{li2016}. 
The quality factor of the bar is defined using the ratio between the imaginary and real parts of the first natural frequency of the free vibration. Figure \ref{fig:QiVsLength} compares the quality factor obtained using current formulation with the Li $et$ $al.$ \cite{li2016} results. Similarly, Figure \ref{fig:QiVsLengthJ} shows a comparison of quality factor of a fixed-fixed bar under different thermal boundary conditions obtained using current formulation with Jiao $et$ $al.$ \cite{jiao2014} results. It can be observed from both figures that the results obtained using current formulation are in good agreement with published results. 

Contrary to flexural vibrations, in case of longitudinal vibration, high-frequency waves (inversely proportional to bar length) are relaxed, and low-frequency waves are un-relaxed in terms of heat transfer across the bar \cite{lifshitz2000}. Maximum dissipation occurs when relaxed and unrelaxed limits meets each other. Inside a relaxed range (at shorter bar lengths), dissipation is higher for adiabatic boundary conditions as compared to isothermal boundary conditions because of higher relaxation time. Similarly, at higher lengths of the bar when limits are un-relaxed, dissipation is higher for isothermal boundary conditions because of shorter relaxation time (Figure \ref{fig:QiVsLengthJ}). According to Zener's theory of thermoelasticity, maximum dissipation occurs when $\omega \tau \approx C$, where $\omega$ is structural eigenfrequency of the bar while $\tau$ is a characteristic time taken to equalize the temperature across the bar length and $C$ is the constant value. 
For a fixed-fixed 1D-bar maximum dissipation occurs at a length ($L_{peak}$) of around 
\begin{flalign}
	\label{Lpeak}
	&L_{peak}\approx C\frac{\pi k_0}{4c_v ^0\sqrt{Y_0 \rho _0}}\mathrm{\,\,\,\,\,\,\,\,for\,adiabatic-adiabatic\, end\, conditions}\\
	\label{Lpeak2}
	&L_{peak}\approx C \frac{\pi k_0}{c_v ^0\sqrt{Y_0 \rho _0}}\mathrm{\,\,\,\,\,\,\,\,for\,isothermal-isothermal\, end\, conditions}
\end{flalign}
Using this mathematical expressions (see Eq. \eqref{Lpeak} and Eq. \eqref{Lpeak2}), $L_{peak}$ for adiabatic-adiabatic boundary conditions and isothermal-isothermal boundary conditions are $35.3$ nm and $159$ nm, respectively. The finite element analysis results (see Figure \ref{fig:QiVsLengthJ}) are in good agreement with the estimated $L_{peak}$ values for both boundary conditions. \\
In the subsequent sections effect of different nonlinearities on the thermoelastic damping of a bar undergoing longitudinal vibrations is discussed.
\subsection{Effect of geometric nonlinearity on the TED in the logitudinal vibration}
\label{gl}  
\begin{figure}
	\centering
	\subfloat[Damping vs length]{
		\includegraphics[width=0.45\textwidth]{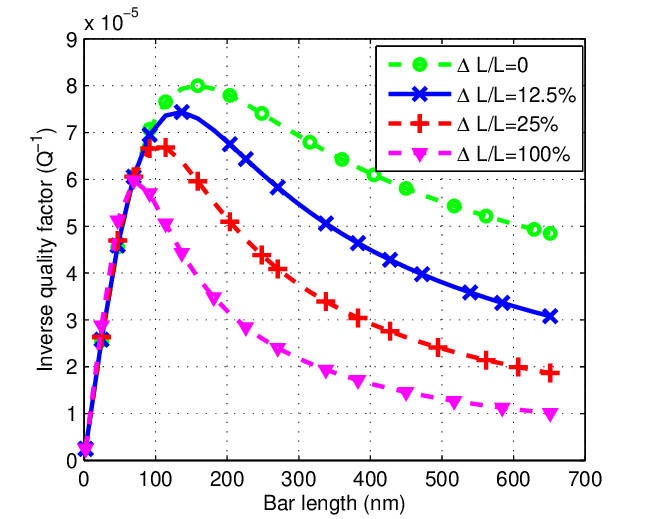}
		\label{DVLGL}}
	\quad
	\subfloat[Frequency shift vs geometric nonlinearity]{
		\includegraphics[width=0.45\textwidth]{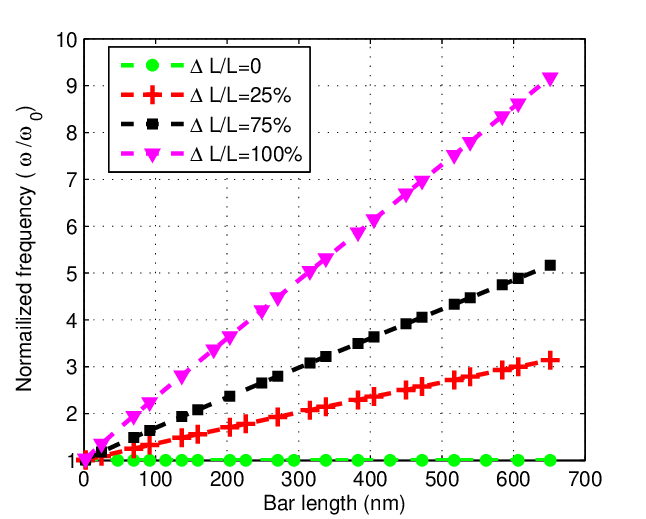}
		\label{FVGL}}
	\caption{Effect of geometric nonlinearity on quality factor and frequency for different degree of geometric nonlinearity for a fixed-fixed bar with isothermal boundary conditions. Normalized frequency shift ($\Delta w/w_0$) is the frequency shift measured with respect to the first natural frequency of the bar of $L=100nm$.}
	\label{QGL}
\end{figure}

The nonlinearity is induced by considering pre-strain in the bar, represented with respect to the original length of the bar.
Figure \ref{QGL} shows the variation of quality factor and the normalized frequency shift ($\Delta \omega/\omega_0$) of a fixed-fixed bar for different degrees ($\Delta L/L$) of geometric nonlinearity.  The normalized frequency shift is the frequency shift measured with respect to the first natural frequency of the bar.
The prestrain in the bar causes increase in the vibration frequency. 
This increase in vibration frequency decreases thermoelastic dissipation at higher bar lengths where the thermoelastic waves are un-relaxed and the opposite effect is seen for the shorter bar lengths.
With pre-strain, the point of maximum dissipation shifts to the left i.e. towards the shorter bar lengths to maintain the constant value of $\tau \omega$.  

\subsection{Effect of internal heat generation on the TED in the logitudinal vibration}
\label{hg} 
\begin{figure}
	\centering
	\subfloat[Damping vs length]{
		\includegraphics[width=0.45\textwidth]{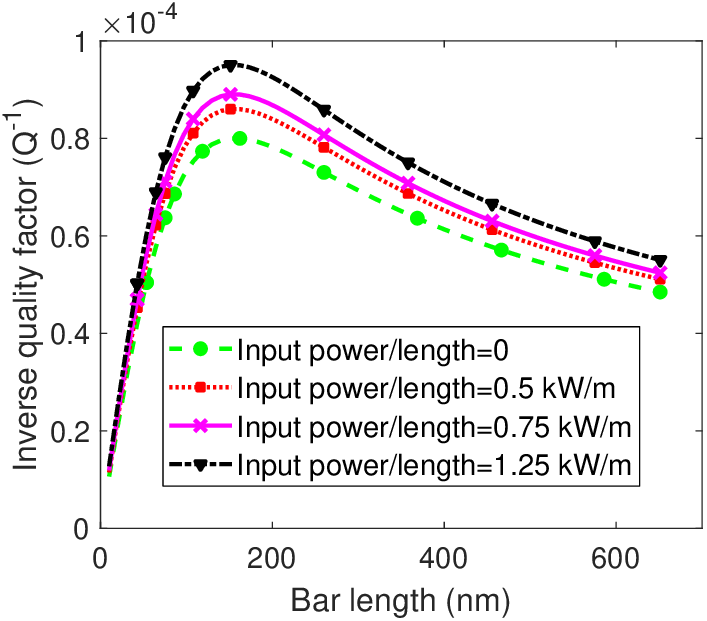}
		\label{QVLHF}}
	\quad
	\subfloat[Frequency shift vs Input power]{
		\includegraphics[width=0.45\textwidth]{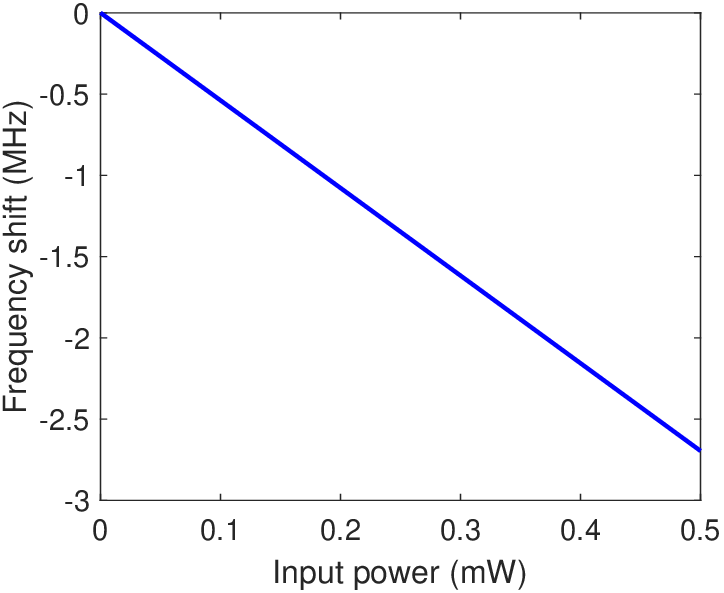}
		\label{FVHF}}
	\caption{Effect of input power/ internal heat generation on quality factor and frequency for a fixed-fixed bar with isothermal boundary conditions. Frequency shift ($\Delta w$) is the shift in frequency from the first natural frequency of a bar of $L=100nm$.}
	\label{QHF}
\end{figure}

Figure \ref{QHF} shows the effect of input power or internal heat generation on the thermoelastic coupling. A large shift in frequency has been observed for a minimal input value of heat. This high sensitivity towards input power can be used to detect small currents by measuring the frequency shift due to Ohmic heating. With increase in the input power or internal heat the thermal relaxation time ($\tau$) increases and at the same time the vibrational frequency $\omega$ decreases. In overall the product $\tau \omega$ is remaining constant. However we see an increase in dissipation over the whole range of bar length due to the increase in the input internal heat (overall temperature of the bar). This observation is inline with the Zener \cite{zener1938} and Zhang \cite{zhang2004} theory for thermal dissipation. In addition, with increase in internal heat generation, the length at which maximum dissipation occurs remains almost unchanged as the product $\tau \omega $ is constant.   

\subsection{Effect of temperature-dependent Young's modulus on the TED in the logitudinal vibration}
\label{etdy}
\begin{figure}
	\centering
	\subfloat[Damping vs length]{
		\includegraphics[width=0.45\textwidth]{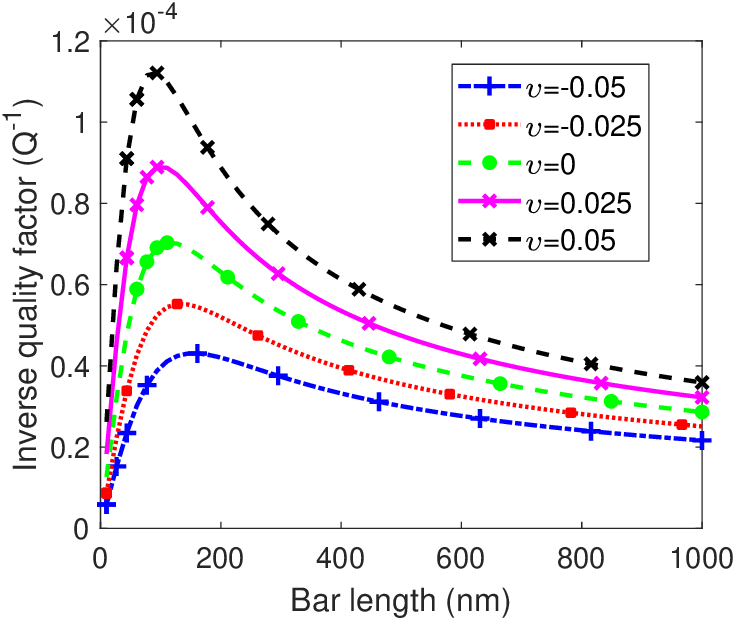}
		\label{QVLY}}
	\quad
	\subfloat[Frequency shift vs length]{
		\includegraphics[width=0.45\textwidth]{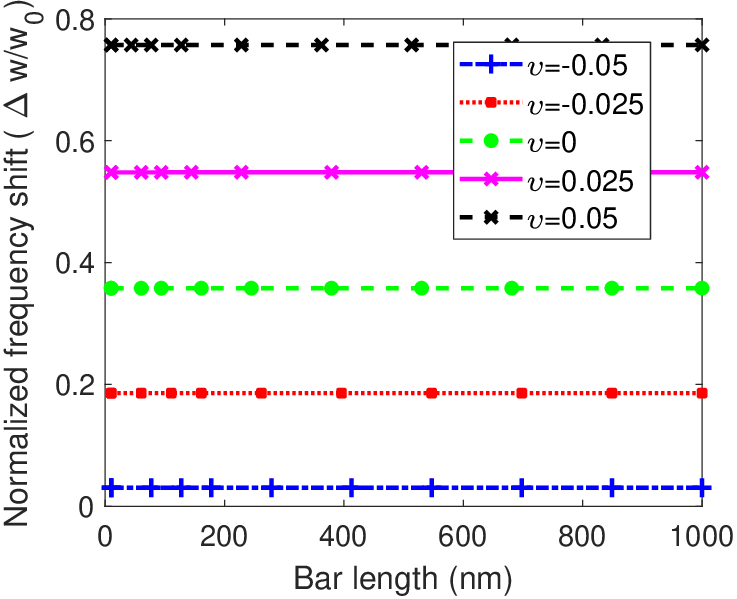}
		\label{FVY}}
	\caption{Effect of temperature dependent Young's modulus on thermoelastic behavior of a fixed-fixed bar with isothermal boundary conditions. The bar has a pre-strain of $\Delta L/L=0.25$ and is under a constant input power of $0.25kW/m$. Young's modulus of the form $Y=Y_0e^{\upsilon (T-T_0)}$}
	\label{QY}
\end{figure}

The effect of temperature dependent Young's modulus is investigated by considering Young's modulus of the form $Y=Y_0e^{\upsilon (T-T_0)}$. Figure \ref{QY} shows the effect of temperature-dependent Young's modulus. As discussed already, a decrease in peak length is observed because of an increase in vibrational frequency to maintain a constant value of $\omega \tau$. The overall behavior of dissipation is in line with Zener \cite{zener1938} and Zhang \cite{zhang2004} theory at a smaller length (when thermal waves are relaxed) where dissipation increases with a decrease in vibrational frequency while its opposite to intuition at higher lengths. This behavior can be explained in terms of strong thermoelastic coupling where the magnitude of damping increases due to an increase in the value of Young's modulus ($\frac{Y_0\alpha _0^2T_0 }{c^0_v}$ \cite{zhang2004}) which overpowers the decrement due to Lorentzian behavior of quality factor with respect to change in frequency.

\subsection{Effect of temperature-dependent thermal conductivity on the TED in the logitudinal vibration}
\label{etdt} 
\begin{figure}
	\centering
	\subfloat[Damping vs length]{
		\includegraphics[width=0.45\textwidth]{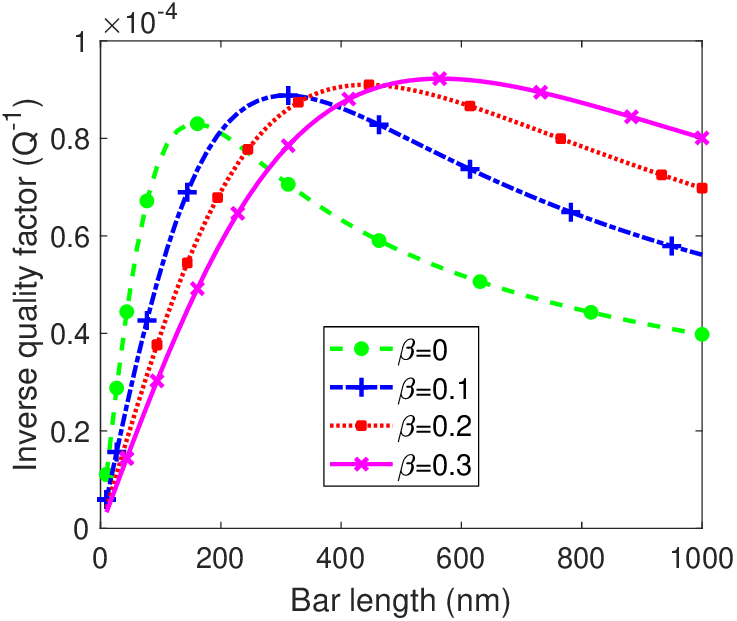}
		\label{QVLC}}
	\quad
	\subfloat[Frequency shift vs length]{
		\includegraphics[width=0.45\textwidth]{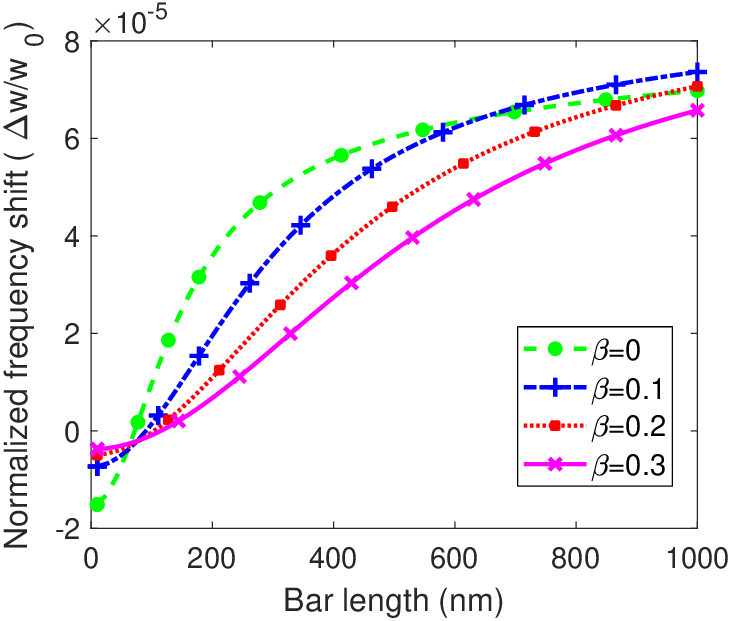}
		\label{FVC}}
	\caption{Effect of temperature dependent coefficient of thermal conductivity on thermoelastic behavior of a fixed-fixed bar with isothermal boundary conditions. Bar is under a constant input power of $0.25kW/m$. Coefficient of thermal conductivity is related to temperature as $k=k_0e^{\beta (T-T_0)}$}
	\label{QC}
\end{figure}

The temperature dependent thermal conductivity of the form $k=k_0e^{\beta (T-T0)}$ is considered to study the effect of it on the TED of a fixed-fixed bar subjected to longitudinal vibrations.
As shown in Figure \ref{QC}, the frequency of the bar decreases with a thermal conductivity parameter $\beta$ because of two-way thermoelastic coupling. However the change in the frequency is not significant. Additionaly, the thermal relaxation time decreases with $\beta$. 
Since the effect of $\beta$ on thermal relaxation time is more prominent as compared to the frequency, dissipation decreases for small lengths (relaxed thermal waves) while it increases for larger lengths (unrelaxed thermal waves). Similarly, the length at which the maximum dissipation occurs increases with increase in thermal conductivity parameter $\beta$  in the view of constant product $\omega \tau$.

\subsection{Effect of strain-dependent thermal conductivity on the TED in the logitudinal vibration}
\label{etdtc} 
\begin{figure}
	\centering
	\subfloat[Damping vs length]{
		\includegraphics[width=0.45\textwidth]{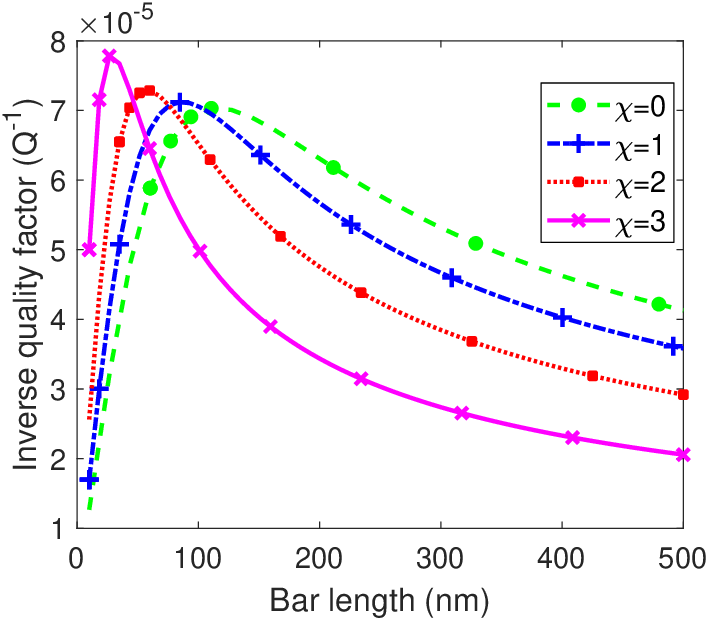}
		\label{QVLSC}}
	\quad
	\subfloat[Frequency shift vs length]{
		\includegraphics[width=0.45\textwidth]{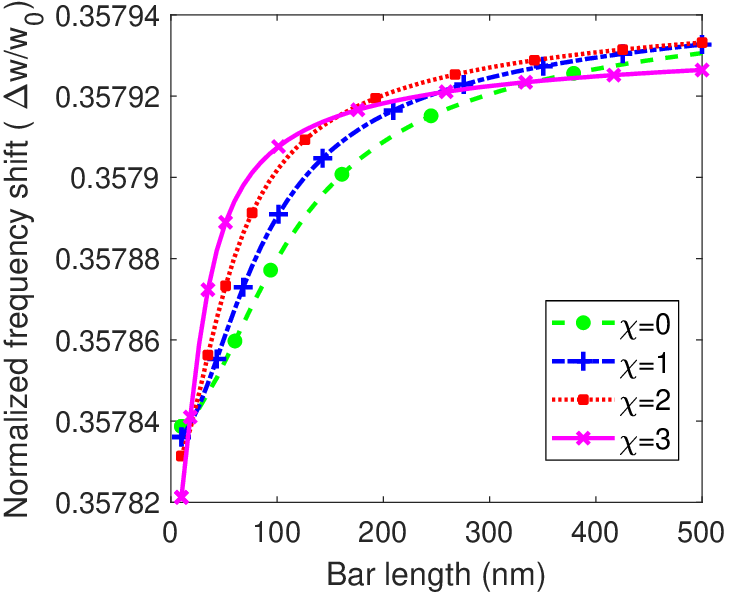}
		\label{FVSC}}
	\caption{Effect of strain dependent coefficient of thermal conductivity on thermoelastic behavior of a fixed-fixed bar with isothermal boundary conditions. The bar has a pre-strain of $\Delta L/L=0.25$ and is under a constant input power of $0.25kW/m$. Coefficient of thermal conductivity is related to strain as $k=k_0\left(1-\chi \frac{\partial u}{\partial x}\right)$}
	\label{QSC}
\end{figure}

To study the effect of strain dependent thermal conductivity on the TED of fixed-fixed bar, a strain dependent coefficient of thermal conductivity is considered as $k=k_0\left(1-\chi \frac{\partial u}{\partial x}\right)$.
Figure \ref{QSC} shows the effect of strain dependent coefficient of thermal conductivity on the quality factor and the frequency shift of bar undergoing longitudinal vibration.
A higher value of $\chi$ decreases thermal conductivity, thus the thermal relaxation time of the bar, which decreases the damping at higher length where waves are un-relaxed. When the length of the bar is less where waves are relaxed, a decrease in thermal conductivity increases the quality factor. The length at which the maximum thermoelastic damping is observed, decreases with increase in value of $\chi$.

\section{Conclusion}
\label{conclusion}
Thermoelastic damping is an important phenomenon seen in the very high frequency MEMS/NEMS resonators. These MEMS/NEMS devices are made of materials whose properties are functions of temperature or strain field. In addition, it is observed that these devices undergo large deformation. This necessitates to include geometric and material nonlinearities while studying thermoelastic behavior. 

In this paper, generalized 3-D finite element formulation of thermoelastic equations is derived, which includes geometric and material nonlinearities in itself. Starting with a brief derivation of thermoelastic equations and constitutive relation, incremental Total Lagrangian equations and  the final nonlinear coupled finite element equations are derived. Next, the current formulation is used to study the thermoelastic damping in the 1-D rods undergoing longitudinal vibrations.

The current formulation is validated by comparing results for 1-D case with COMSOL Multiphysics simulation (for load-displacement characteristics in large deformation regime), analytical results ( for material nonlinearity) and published results (for linear thermoelastic coupling).

Thermoelastic damping in a longitudinally vibrating 1-D bar is analyzed by studying the variation of quality factor and frequency with the length of the bar for different combinations of nonlinearities. It was found that in such problems thermal and strain field is strongly coupled; for example, large strain affects the thermal relaxation time of the bar while large temperature variation affects natural frequency in the bar quite significantly. This formulation can be used to analyze 2-D and 3-D thermoelastic problems such as thermoelastic damping in plates and flexural beams.
\bibliography{Full}
\end{document}